\RequirePackage[2020-02-02]{latexrelease}
\documentclass[preprint,12pt]{revtex4}

\usepackage{braket}
\usepackage{graphicx}
\usepackage[printfigures]{figcaps}
\usepackage{amsmath,amssymb}

\begin{document}
\title{Mechanism of magnetic diode in artificial honeycomb lattice}
\author{G. Yumnam$^{1}$}
\author{J. Guo$^{1,\dagger}$}
\author{Y. Chen$^{1}$}
\author{A. Dahal$^{1}$}
\author{P. Ghosh$^{1}$}
\author{Q. Cunningham$^{1}$}
\author{J. Keum$^{2}$}
\author{V. Lauter$^{2}$}
\author{A. Abdullah$^{3}$}
\author{M. Almasri$^{3}$}
\author{D. K. Singh$^{1,*}$}
\affiliation{$^{1}$Department of Physics and Astronomy, University of Missouri, Columbia, MO 65211}
\affiliation{$^{2}$Oak Ridge National Laboratory, Oak Ridge, TN 37831}
\affiliation{$^{3}$Department of Electrical Engineering, University of Missouri, Columbia, MO 65211}
\affiliation{$^{\dagger}$Author made equal contribution}
\affiliation{$^{*}$Email: singhdk@missouri.edu}

\begin{abstract}
Spin diode is important prerequisite to practical manifestation of spin electronics. Yet, a functioning magnetic diode at room temperature is still illusive. Here, we reveal diode-type phenomena due to magnetic charge mediated conduction in artificial honeycomb geometry, made of concave shape single domain permalloy element. We find that honeycomb lattice defies symmetry by populating vertices with low and high multiplicity magnetic charges, causing asymmetric magnetization, in applied current of opposite polarity. High multiplicity units create highly resistive network, thereby inhibiting magnetic charge dynamics propelled electrical conduction. However, practical realization of this effect requires modest demagnetization factor in constituting element. Concave structure fulfills the condition. Subsequently, magnetic diode behavior emerges across broad thermal range of $T$ = 40K - 300K. The finding is a departure from the prevailing notion of spin-charge interaction as the sole guiding principle behind spintronics. Consequently, a new vista, mediated by magnetic charge interaction, is envisaged for spintronic research.
\end{abstract}

\maketitle

Spin or magnetic diode is at the center of spin-based logic operation and the scalability of spin electronics.\cite{Zutic,Tulapurkar,Kim} Current efforts in the exploration of magnetic diode are focused on utilizing the spin-charge interaction mechanism in multilayered magnetic thin films and semiconducting materials, such as (GaMn)As.\cite{Kent,Chappert,Flatte2,Dorpe} The study has not only led to the development of bipolar and unipolar prototypes with very high level of unidirectional spin polarization,\cite{Vignale,Acremann,Ouna} but has also resulted in better theoretical understanding of the spin torque mechanism in spin transport phenomena.\cite{Zutic2} Even though the synergistic research approaches have made remarkable progress in the pursuit of spin-based rectifier, their application is limited due to two reasons:\cite{Flatte2,Molenkamp} first, the phenomenon is often limited to low temperature and second, magnetic field application is necessary to achieve unidirectional conduction. A new research venue is needed to spur the realization of a magnetic diode, which could function at room temperature without magnetic field application. Here, we show that the emergent physics of magnetic charge correlation, realized in artificial spin ice magnet under the dumbbell prescription of magnetic moment, provides a new formalism to the exploration of energetically efficient magnetic diode. Magnetic charges are quasiparticles, represented by the quantum-mechanical Pauli matrices in Hamiltonian formulation.\cite{gei} In a highly unusual observation, we find that the indirect interaction between magnetic charges and electric charge carriers induces asymmetric electrical conduction in thermally tunable honeycomb lattice with modest demagnetization energy. The occurrence of robust diode-type unidirectional electrical transport at room temperature signifies the role of emergent quasiparticle in elucidating spintronics property in nanostructured spin ice material.

\begin{figure*}
\centering
\includegraphics[width=18 cm]{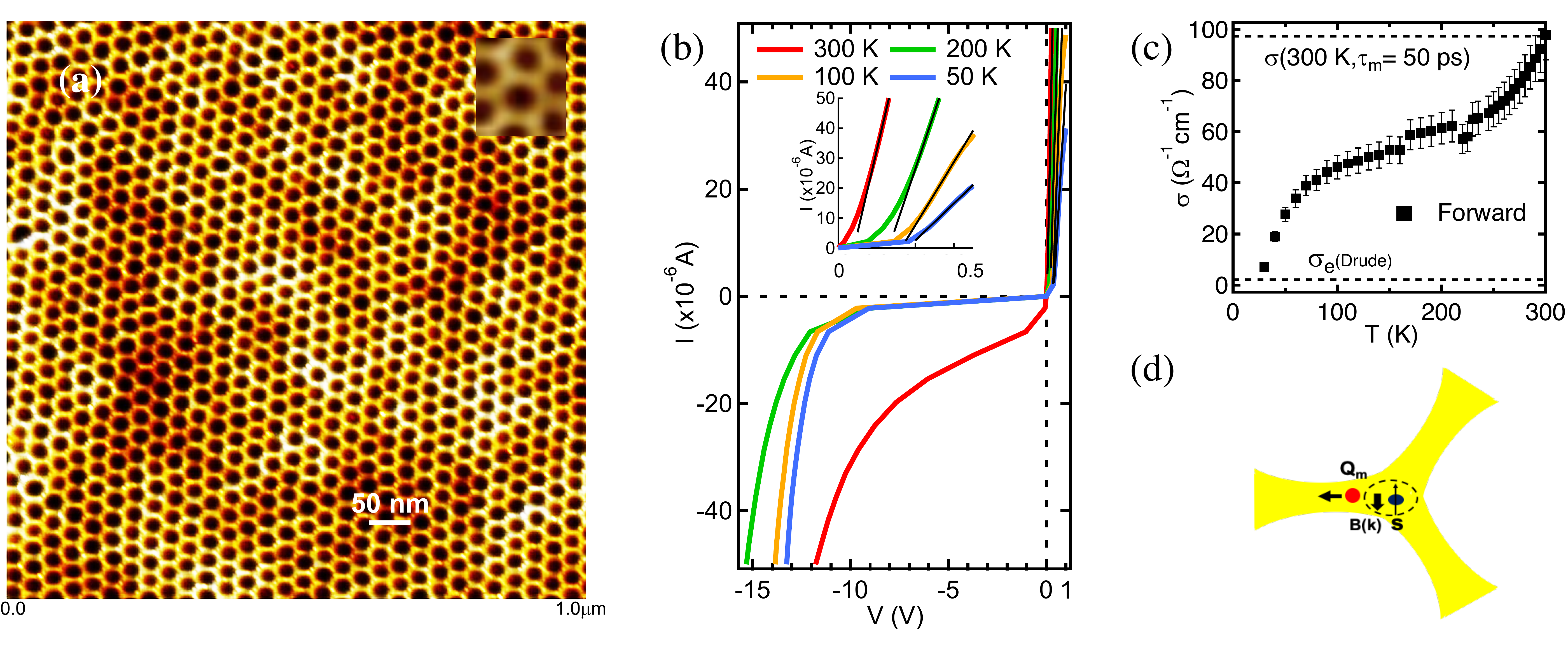} \vspace{-4mm}
\caption{Diode-type asymmetric electrical transport and temperature dependence of conductivity. (a) Atomic force micrograph of artificial honeycomb lattice of concave shape single domain size elements (used in this study). (b) I-V traces at few characteristic temperatures. Electrical conductivity ($\sigma$) is extracted via linear fit to the data. (c) Plot of $\sigma$ vs. $T$(K). The lower dashed line indicates estimated background conductivity due to purely electron's relaxation, while the upper dashed line is estimated using magnetic charge mediated conduction phenomenology. In forward bias, the conductivity increases by more than an order of magnitude at $T$ = 300 K. (d) Figure shows the schematic of magnetic charge mediated conduction in honeycomb element (see text for detail).} \vspace{-4mm}
\end{figure*}

The system we study is a recently realized frustrated artificial honeycomb, made of concave shape single domain size permalloy (Ni$_{0.81}$Fe$_{0.19}$) elements, shown in Fig. 1a (see Experimental Section for nanofabrication details).\cite{Glavic} The single domain size magnetic element ($\sim$ 12 nm in length ($L$)) makes it a thermally accessible system with the inter-elemental dipolar interaction energy of $\sim$15 K.\cite{Glavic} Magnetic honeycomb lattice manifests natural tendency to develop magnetic charges on its vertices due to the ice-rule.\cite{Nisoli,Stamps} A magnetic moment is considered as a dumbbell of '+q' and '-q' magnetic charges on the ends that interact via magnetic Coulomb's interaction.\cite{Sondhi,Bramwell,N} Therefore, net magnetic charge Q on a honeycomb vertex is given by the algebraic summation of all individual '+q' and '-q' charges e.g. Q = $\sum q_i$ where q$_{i}$ = M/$L$, M being the net magnetic moment aligned along the length of connecting element due to shape anisotropy.\cite{Nisoli,Stamps} In a honeycomb, magnetic charges occur in two integer multiplicities of $\pm$Q and $\pm$3Q units-- associated to the local magnetic moment arrangements of two-in \& one-out (or vice-versa) or, all-in or all-out configurations, respectively.\cite{Stamps} The magnetic charge pattern in thermally tunable lattice is constantly seeking a low entropic configuration by emitting or absorbing dynamic magnetic charge defect, Q$_{m}$, of magnitude $\mid$2Q$\mid$ unit. The dynamic nature of magnetic charge defect was recently evidenced from neutron spin echo measurements on permalloy honeycomb sample.\cite{Chen2}

Thermally tunable magnetic honeycomb lattice manifests distinct diode-type asymmetric conduction in electrical measurement. In Fig. 1b, we show the plot of I-V traces, obtained using the two-probe measurement technique (see Experimental Section for detail), at few characteristic temperatures. The diode behavior in honeycomb sample is also verified using the four probe electrical measurement technique (see Fig. S1 in Supporting Information). The unidirectional conduction persists across the entire measurement range of $T$ = 30 K to $T$ = 300 K. Furthermore, the system requires a smaller threshold voltage (compared to the conventional semiconductor diode), $\sim$ 50 mV, to drive significant current in forward bias direction at room temperature. The estimated conductivity in the forward biased state is plotted as a function of temperature in Fig. 1c. The forward conductivity depicts non-monotonous increase in temperature with respect to the purely electrical background of $\sigma$$_{0}$ $\sim$ 3 $\Omega$.cm. The background conductivity is estimated using the Drude's formula ($n$e$^{2}$$\tau_e$/m$_{e}$) by utilizing the previously reported values of electrons density ($n$) $\sim$ 10$^{21}$ cm$^{-2}$ and electric charge relaxation time ($\tau_e$) $\sim$10$^{-14}$ s in permalloy thin film.\cite{Bansil,Petrovykh} 

The additional conductivity in the forward biased state is attributed to the indirect interaction between magnetic charge defect dynamics and electric charge carriers. Magnetic charge defect was originally conceived to explain the magnetic monopoles in bulk spin ice.\cite{Nisoli,Sondhi} The charge defect's dynamics in spin ice generates transverse fluctuations in local magnetic field $\textbf{B(k)}$ that couple with conduction electron's spin \textbf{s} to spur magnetic charge mediated conduction.\cite{Chern} We test the applicability of the hypothesis in artificial honeycomb, as narrow constriction, $\sim$ 5 nm, and small thickness, $\sim$8 nm, of connecting element restricts the charge defect's motion along its length, causing transverse fluctuation in local magnetic field perpendicular to the direction of motion. It is schematically described in Fig. 1d. According to the formalism,\cite{Chern} electrical conductivity depicts a temperature dependence of $\sigma$ = $\sigma_0 \varepsilon_F \tau_m$/($\varepsilon_F \tau_m - \alpha h l_c k_F .e^{-\Delta/k_BT}$), where $\epsilon_{F}$ and $k_F$ are Fermi surface energy and wavevector, $l_c$ is the cut-off length for transverse fluctuation in magnetic field, limited by the thickness of the honeycomb film, $\alpha$ is a temperature independent constant due to magnetic lattice and $\Delta$ is the threshold energy to create magnetic charge defect, given by the energy difference of $\mid$E$_{3Q}$ - E$_{Q}$$\mid$ or, $\mid$E$_{Q}$ - E$_{-Q}$$\mid$ $\sim$ 30 K. $\tau_m$, the relaxation time of quasi-particle, is a key parameter in conductivity formulation. The value of $\tau_m$ in thermally tunable honeycomb lattice $\sim$ 50 ps,\cite{Chen2} as inferred from neutron spin echo measurements (see Fig. S2), is comparable to the spin relaxation time in spin ice material.\cite{Ehlers} It further supports the applicability of the model in the present case. As shown in Fig. 1c, the estimated conductivity in the forward biased state is in good agreement with the calculated $\sigma$ at $T$ = 300 K due to the magnetic charge mediation mechanism.

\begin{figure*}
\centering
\includegraphics[width=18 cm]{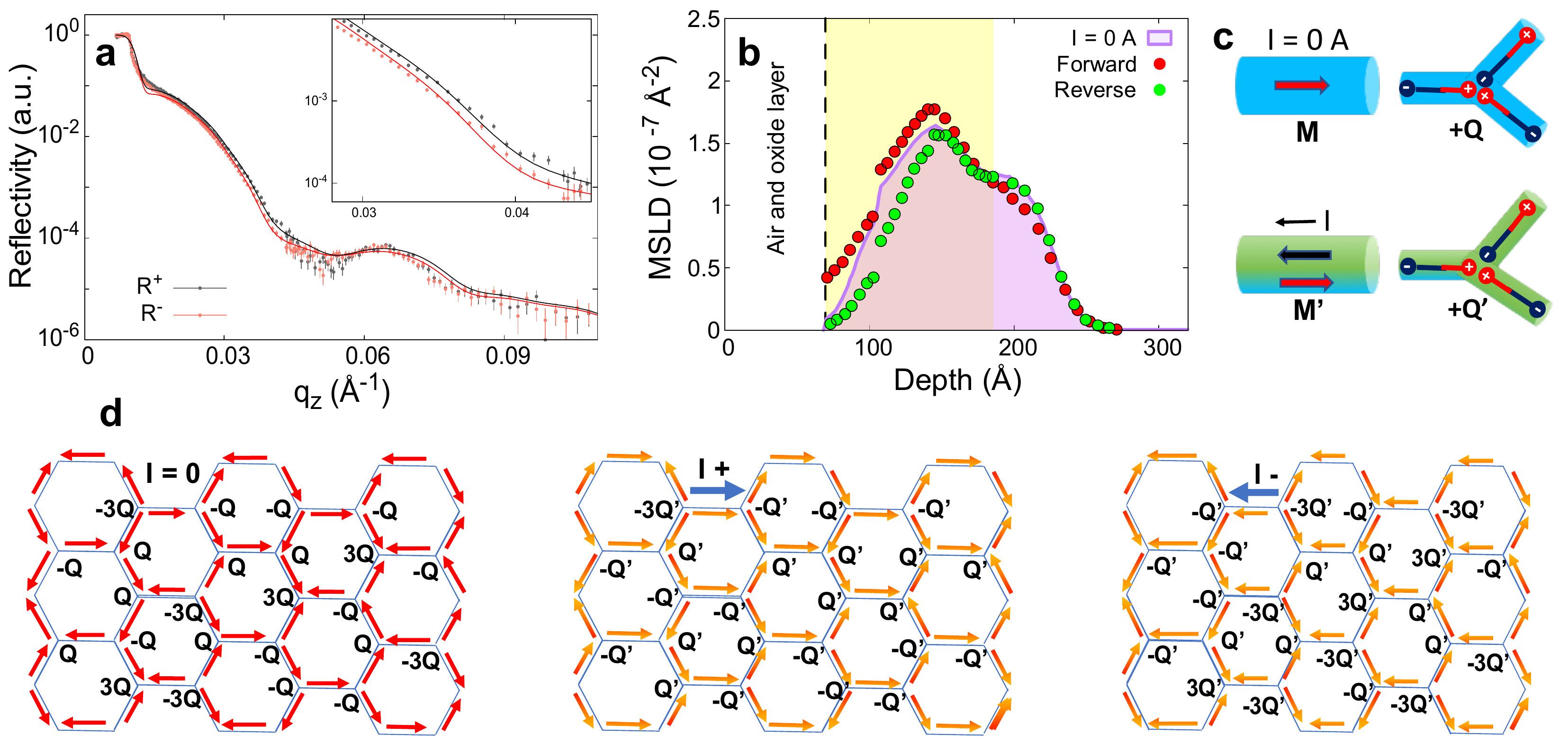} \vspace{-2mm}
\caption{Current induced fractional transformation of magnetic charge density in permalloy honeycomb. (a) Polarized neutron specular reflectivity data as a function of wave vector in unbiased state (I = 0 A) at $T$ = 300 K (also see Fig. S2 for the plot of specular data in current biased states in SI). A small guide field of $H$ = 20 Oe is applied to keep neutron polarized. Error bar represents one standard deviation. (b) Plot of magnetic layer thickness (related to bulk magnetization) as a function of the lattice thickness, obtained from fitting the specular PNR data. Current biasing of 25 $\mu$A induces partial modification to magnetic moment direction, starting from the top layer, along the current application direction. Overall, the magnetization increases (decreases) in forward (reverse) biased states, compared to the unbiased state. (c) Schematic description of partial moment flipping. (d) Magnetic charge pattern across honeycomb vertices in unbiased state (consisting of random distribution of $\pm$Q and $\pm$3Q charges), forward biased state (consisting of mostly fractionally transformed $\pm$Q' charges) and reverse biased state (consists of mostly $\pm$3Q'). Current application only flips moments that were strictly aligned against current propagation in the original unbiased state. } \vspace{-4mm}
\end{figure*}

The magnetic charge mediated electrical conduction is suppressed in opposite current bias direction. To understand the connection between diode-type electrical transport and magnetic charge evolution in the current biased states, we have performed spin polarized neutron reflectometry (PNR) measurements on honeycomb sample in multi-variable sample environment of current, temperature and magnetic field. Unlike an external tuning parameter, such as magnetic field, which is energetic enough to reverse the magnetization direction along a honeycomb element, PNR measurements reveal that the current application only partially modifies the overall moment. PNR measurements were performed at $T$ = 300 K in the unbiased and $\pm$ 25 $\mu$A current biased states (see Experimental Section). The specular reflectivity plot in the unbiased state is shown in Fig. 2a. Irreversible spin polarized neutron reflectivities, R+ and R- ('+' and '-' represent 'spin up' and 'spin down' neutron components), indicate magnetic nature of scattering. Similar behavior is observed in the current biased states (see Fig. S3 for specular data in current biased states). Analysis of experimental data using the generic LICORNE program yields important information about the magnetic layer thickness and the macroscopic magnetization.\cite{Lauter} We show the plot of magnetic depth profile in zero and $\pm$ 25 $\mu$A current biases in Fig. 2b. The depth profile reveals two interlinked extraordinary current bias effects: first, the overall magnetization increases (decreases) in forward (reverse) current bias configuration. Second, only a partial flipping of moment along the honeycomb element takes place in the current biased state. Current bias tends to align magnetic moment, originally pointed against the flow, to the current application direction. However, it is not strong enough to flip the entire moment. Rather, magnetic moment inversion process is limited to a partial thickness, about 60\%, only; starting from the top layer of the honeycomb film. Consecutively, magnetic moment of individual element changes from M, in the unbiased state, to M' = (0.6 - 0.4) M = 0.2 M in the current biased state, but now aligns along the current bias (schematically described in Fig. 2c). Moments that were originally aligned to the current bias direction, remain unaffected. The effect of current-induced magnetic moment inversion along honeycomb element can increase or decrease macroscopic magnetization. In an artificial magnetic honeycomb lattice, only $\pm$Q charges give rise to the finite magnetization due to the short-range correlation of '2-in \& 1-out' (or vice-versa) moment arrangement. Honeycomb vertices with $\pm$3Q charges do not contribute to the net magnetization. Therefore, the enhanced magnetization in forward current bias state must be accompanied by the increased population density of low multiplicity $\pm$Q charges. Similarly, a decrease in magnetization suggests the higher density of $\pm$3Q charges in the reverse biased state. But, magnetic charges no longer have the same magnitude. The reduced moment size, from M to M', in the current biased state introduces a fractional correction to the unbiased magnetic charge, i.e. an individual charge q$_{i}$, corresponding to M/$L$, transforms into the smaller charge q$_{i}$', defined by M'/$L$, in the current biased state. Consequently, charges on the honeycomb vertices undergo transformation from $\sum q_i$ (= $\pm$Q or $\pm$3Q) to $\sum$ m$q_i$'+n$q_i$ (= $\pm$Q' or $\pm$3Q') $\forall$ $m, n$ = 0,1,2,3 and $m+n$ = 3, assuming that the magnetic moment along at least one element is partially flipped. 

\begin{figure*}
\centering
\includegraphics[width=18 cm]{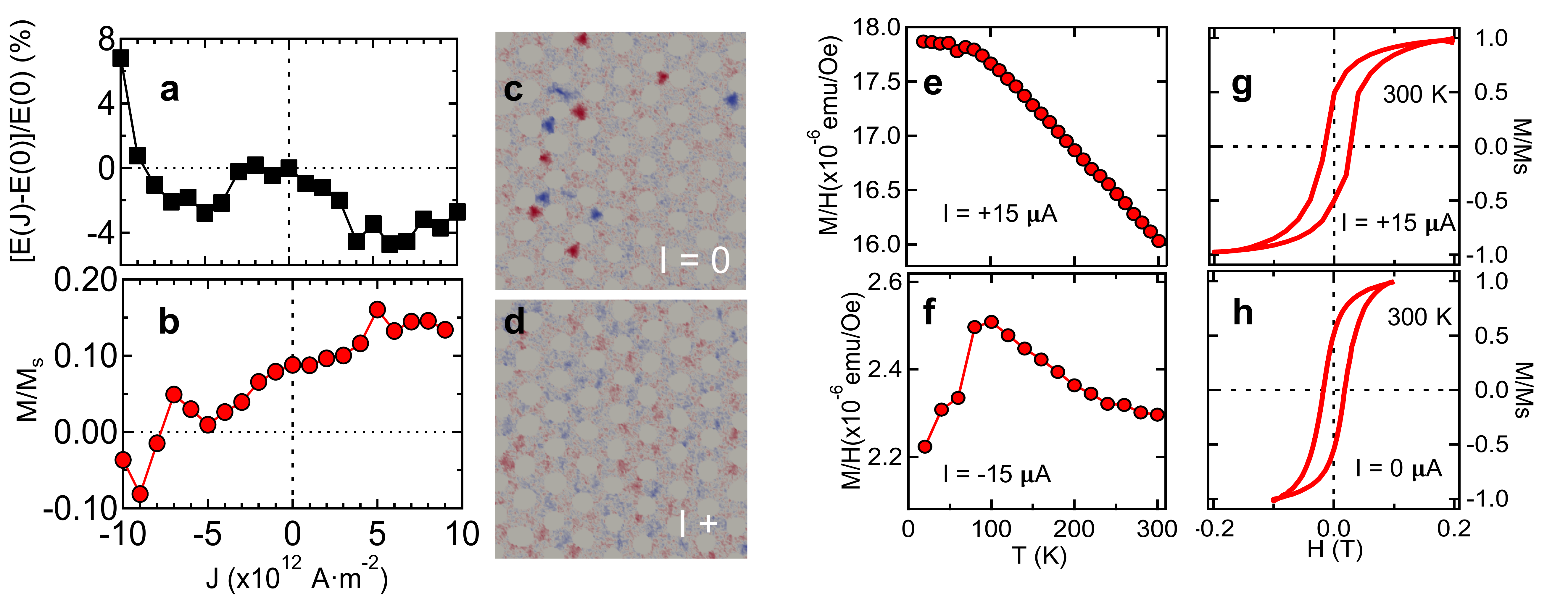} \vspace{-4mm}
\caption{Evolution of asymmetric energy as a function of current and magnetic measurements. (a-b) Micromagnetic simulations at $T$ = 300 K reveal that magnetization increases and energy decreases in forward biased state, compared to the unbiased case. (c-d) Simulated magnetization profiles show that the unbiased state consists of a random distribution of $\pm$Q (light blue and red spots) and $\pm$3Q charges (dark blue and red spots). Dark blue and red spots (indicative of $\pm$3Q charges) disappear in forward biased state, making the lattice comprised of $\pm$Q' charges. (e-f) Magnetization as a function of temperature, measured in $H$ = 25 Oe field, in I = $\pm$15 $\mu$A, respectively. (g-h) Magnetic hysteresis measurements at $T$ = 300 K in I = 15 $\mu$A and 0 A, respectively.} \vspace{-4mm}
\end{figure*}

In the unbiased state at $T$ = 300 K, $\pm$Q and $\pm$3Q charges are expected to be randomly distributed on honeycomb vertices. For a random distribution of magnetic charges, the current induced evolution to plausible charge states is shown in Fig. 2d. In a remarkable observation, we find that the forward (reverse) biasing inadvertently causes higher population density of $\pm$Q' ($\pm$3Q') charges, compared to the unbiased state. The symmetry breaking phenomena seems to be independent of the random charge configuration but depends on current application direction with respect to the honeycomb axes (see Fig. S4 for more example). Correspondingly, a higher (lower) net magnetization is expected in the forward (reverse) biased state. The non-trivial micromagnetic (MM) simulations in the current biased state independently confirm this effect, while also providing new information. MM simulations were performed on MUMAX platform at $T$= 300 K under the pretext of the Landau-Lifshitz-Gilbert equation. Current and thermal fluctuation perturbations to the random state are introduced using the spin-torque and Langevin's dynamics, respectively.\cite{MM,Zhang} Hence, the Hamiltonian is given by,
\begin{equation}
\begin{aligned}
\frac{d\mathbf{m}}{dt}= &\frac{\gamma_{LL}}{1+\alpha^2}\left(\mathbf{m}\times{\mathbf{B}}_{eff}+\alpha\left(\mathbf{m}\times\left(\mathbf{m}\times{\mathbf{B}}_{eff}\right)\right)\right) 	\\
& + \frac{1}{1+\alpha^2} \left(1+\beta\alpha\right)\mathbf{m}\times\left(\mathbf{m}\times\left(\boldsymbol{\upsilon}\cdot\nabla\right)\mathbf{m}\right)   \\
& + \left(\beta-\alpha\right)\mathbf{m}\times\left(\boldsymbol{\upsilon}\cdot\nabla\right)\mathbf{m}  
\end{aligned}
\end{equation}
where $M_s$, $\textbf{m}$, $\gamma_{LL}$, $\alpha$ and $\beta$ are saturation magnetization, current or temperature dependent magnetization, gyromagnetic ratio, damping parameter and the degree of non-adiabaticity, respectively. Effective magnetic field \textbf{B}$_{eff}$ consists of \textbf{B}$_{ex}$ + \textbf{B}$_{magnetostatic}$ + \textbf{B}$_{anisotropy}$ + \textbf{B}$_{thermal}$. The current term $\boldsymbol{\upsilon}$ is given by, $\boldsymbol{\upsilon}$ = ($\mu_B \mu_0$/2e$\gamma_0 M_{sat}$(1+$\beta^{2}$))$\mathbf{j}$. A MUMAX initiated random magnetization profile, relaxed for 8 ns, is selected in the unbiased state. The simulated plots of total energy and magnetization vs. current in zero magnetic field are shown in Fig. 3a and 3b, respectively. Here we see that both magnetization and energy manifest highly asymmetric responses to the current application. While the forward biasing expectedly enhances magnetization and reduces energy ($\pm$ Q charges have lower energy than $\pm$3Q charges), negative current bias has opposite effect. Magnetic charge configuration in the forward biased state seems to primarily consist of $\pm$Q charges, as shown in FIg. 3d. Unlike the asymmetric current bias effect in our honeycomb lattice, the current effect is symmetric in honeycomb made of uniform rectangular shaped elements of similar size (see Fig. S5); typically achieved in samples created using the electron-beam lithography technique.\cite{Nisoli,Stamps} The dichotomy in results can be understood by considering the contribution of various energy terms to the total energy. We show the plots of important energy terms, such as anisotropy, exchange, demagnetization as well as total energy, as a function of current bias in Fig. S6. As we can see, the demagnetization energy decreases significantly in applied current in concave shape element of honeycomb lattice. On the other hand, the rectangular element manifests stronger demagnetization, which resists subtle current effect. We infer that weaker demagnetization facilitates current induced asymmetric magnetization in concave element honeycomb, which is reflected in the asymmetric current responses to anisotropy and exchange terms. The overall magnetization remains mostly unaffected to current application in the rectangular element honeycomb. 

\begin{figure*}
\centering
\includegraphics[width=18 cm]{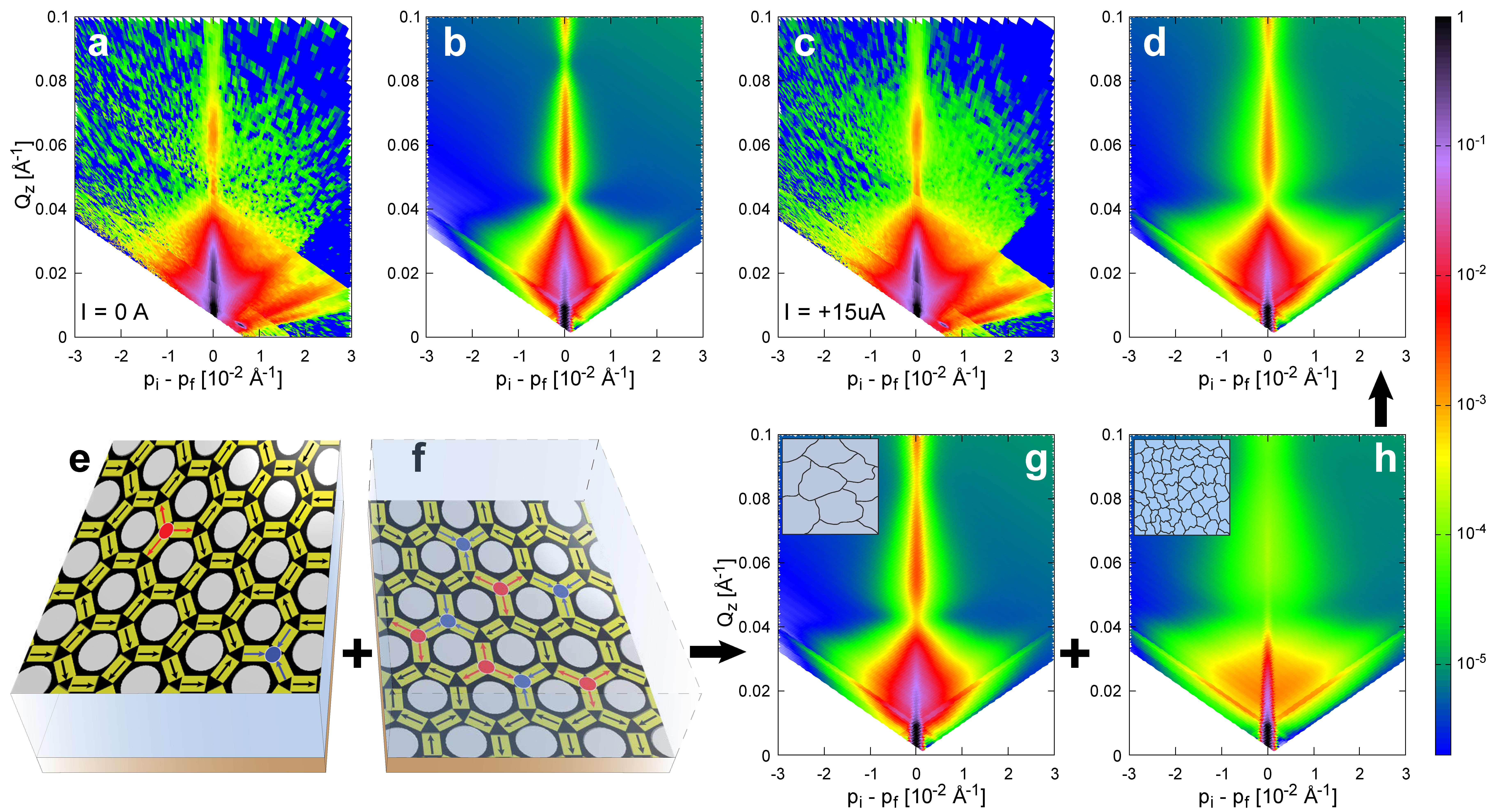} \vspace{-3mm}
\caption{Evidence to proposed magnetic charge patterns in unbiased and current biased states. (a, c) Off-specular reflectometry plots in zero and forward biased states (Fig. c). Polarized neutron data are shown as sum of spin-up and spin-down components. Here, the y-axis represents the out-of-plane scattering vector ($Q_z$= $\frac{2\pi}{\lambda} (\sin{\alpha_i} + \sin{\alpha_f})$), while the difference between the z-components of the incident and the outgoing wave vectors ($p_i - p_f$ = $\frac{2\pi}{\lambda}(\sin{\alpha_i} - \sin{\alpha_f})$) is drawn along the x-axis. (b, d) Numerically simulated reflectometry patterns for charge configurations in unbiased and forward biased states shown in Fig. 3d. (e-h) DWBA simulation for current biased state involves magnetic charge transformation due to partial flipping of magnetic moment. While the bottom layer still maintains unbiased charge configuration (Fig. f), top layer (Fig. e) manifests fractionally transformed magnetic charge pattern. DWBA simulations also take into account structural domains in honeycomb sample. Fig. g and Fig. h are simulation results for different structural domains of 4 $\mu$m and 0.8$\mu$m (shown in the insets), respectively. Domain averaged simulated profile in current biased state (Fig. d) explains experimental data.} \vspace{-6mm}
\end{figure*}

We have also performed magnetic measurements in applied current that are in agreement with the PNR measurements and the micromagnetic simulations. As shown in Fig. 3e-f, the distinctly different magnetization responses are detected in the forward and the reverse biased states. In the forward biased state, the overall magnetization tends to increase as temperature decreases, suggesting the expansion of short-range ordered '2-in \& 1-out' or vice-versa magnetic configuration in the lattice. On the other hand, magnetization decreases in the reverse biased state after registering a modest gain above $T$  $\sim$ 100 K, which can be attributed to the thermal fluctuation effect. More importantly, we observe a large magnetization difference between the forward and the reverse biased states at $T$ = 300 K-- consistent with both the PNR data and the MM simulation. Magnetic hysteresis loops in the unbiased and the current biased states do not reveal any major difference, see Fig. 3g-h. Magnetization tends to saturate above $H$ = 500 Oe in both the unbiased and the biased states, which suggests that the magnetic charge dynamics is ceased to exist in large field application. Indeed, a magnetic field application of 500 Oe or higher suppresses the magnetic charge mediated conduction in the forward biased state. It is prominently depicted in the differential conductivity measurements, see Fig. S1b.\cite{Summers}  

The direct confirmations to the proposed charge configurations in Fig. 2d are obtained from the modeling of the off-specular PNR data. In Fig. 4a, c, we show the plots of off-specular data in zero and applied current states. The specular reflectivity lies along the x = 0 line. Unlike in the unbiased state where the spectral weight is confined to the specular reflection, noticeable off-specular scattering along Q$_{x}$ is observed between Q$_{z}$ = 0.04 $\AA^{-1}$ and 0.08 $\AA^{-1}$ in the forward biased state. A broad feature along the horizontal axis usually indicates the development of inplane magnetic correlation. To elucidate the nature of magnetic charge arrangement on honeycomb vertices, we model the data using Distorted Wave Born Approximation (DWBA) on BornAgain platform (see details in Supporting Information).\cite{Glavic} In the modeled charge structure, the bottom layer of the honeycomb element still maintains the unbiased charge configuration (as described schematically in Fig. 2c), while magnetization in the top layer is flipped along the current application direction. The simulation was domain averaged to account for the structural domains in the honeycomb sample. As shown in Fig. 4, the simulated reflectometry profile reproduces essential features of the experimental data. DWBA modeling vindicates the current-induced fractional transformation of magnetic charges, as shown in Fig. 2c-d. The delicate effect of current induced fractional correction to magnetic charge density is overwhelmed by the more energetic magnetic field parameter (see Fig. S7).

The dominance of $\pm$Q' charges in the forward biased state is more favorable for electrical conduction, compared to $\pm$3Q' charges in the reverse biased state. A comprehensive analytical calculation to this effect, confirming the finding, was previously reported in artificial magnetic honeycomb lattice.\cite{Chen} For calculation purposes, a Kagome network is constructed by bisecting the connecting elements. As shown in the Supporting Information, each Kagome side can have one of the two resistances, associated to 'one-in \& one-out' ($R$') and 'both-in' magnetic moments ($R$). Correspondingly, a honeycomb lattice with only $\pm$3Q-type charges will transform into a Kagome resistor network of the same-type of resistance, R. On the other hand, the Kagome network will consist of both $R$ and $R$' resistances in the case of $\pm$Q-type charges on honeycomb vertices, see Fig. S8. The calculation estimates total resistance of the network in each cases in terms of resistances $R$ and $R$'. Now, if we make a justifiable assumption, within the framework of the drift-diffusion model,\cite{Vignale2} that $R$' is smaller than $R$ (noting that 'one-in \& one-out' and 'both-in' represent parallel and anti-parallel type of moment arrangements, respectively), then net $R$$_{Q,-Q}$ is several times smaller than $R$$_{3Q,-3Q}$. Even for a very modest ratio of 1/3 between $R$' and $R$, $R$$_{3Q,-3Q}$ is significant larger than $R$$_{Q,-Q}$ (see Table S1). It suggests that $\pm$3Q-type charges suppress magnetic charge mediated conduction in the reverse biased state by inhibiting electrical transport. It is the final link to the explanation of magnetic diode phenomenon in permalloy honeycomb lattice.

The concept of magnetic charges was originally envisaged to study a fundamental problem of the emergence of effective magnetic monopoles in spin ice materials.\cite{Sondhi,Bramwell} Since then a plethora of research activities has taken place, especially in artificial spin ice where the implication of magnetic charge correlation in magnetic monopoles avalanche and the chiral spin vortex loop ordering is vividly demonstrated.\cite{Mengotti,Glavic} Here, we have elucidated a practical aspect of magnetic charge driven physics, which was not known before. We have shown that the cumulative effect of magnetic charge dynamics and geometrical shape of the constituting element causes an unusual diode-type asymmetric conductivity in magnetic honeycomb lattice. The honeycomb structure, which violates mirror symmetry, plays crucial role in generating the asymmetric charge distributions in the current biased states that are directly linked to the diode effect. In principle, this phenomenon can be realized in any thermally tunable honeycomb lattice with modest demagnetization energy. An attractive aspect of magnetic diode phenomena is a low threshold voltage to drive significant amount of current in the forward biased state. Consequently, a very low power dissipation rectifier emerges. The demonstrated functionality is envisaged to spur new spintronic applications.

\section*{Experimental methods}

\subsection*{Nanofabrication of artificial permalloy honeycomb lattice} Fabrication of artificial magnetic honeycomb lattice starts with the synthesis of porous hexagonal diblock template on top of a silicon substrate. The template fabrication process utilizes diblock copolymer polystyrene(PS)-b-poly-4-vinyl pyridine (P4VP) of molecular weight 29k Dalton with the volume fraction of 70\% PS and 30\% P4VP. The diblock copolymer tends to self-assemble, under right condition, in a hexagonal cylindrical structure of P4VP in the matrix of polystyrene. Submerging the samples in ethanol for 20 minutes releases the P4VP cylinders yielding a porous hexagonal template. Reactive ion etching with CF$_4$ gas was performed to transfer the hexagonal pattern, including the concave shape of the connecting element, to the underlying silicon substrate. RIE was performed at 50 W power at 100 mTorr CF$_{4}$ gas pressure for 20 seconds. The top layer of the substrate resembles a honeycomb pattern. This topographical property is exploited to create magnetic honeycomb lattice by depositing $\sim$ 8 nm thick permalloy on top of the uniformly rotating substrate in near parallel configuration. It produces the desired magnetic honeycomb lattice with a typical element size of 12 nm (length)$\times$~5 nm (width)$\times$8 nm (thickness). Atomic force micrograph of the artificial honeycomb lattice, used in this study, is shown in Fig. 1a. 

\subsection*{Electrical measurements} Electrical measurements were performed in both the two-probe and the four-probe configurations using a synchronized combination of the Keithley current source meter 6221 and a nanovoltmeter 2182A via a trigger link. While the two-probe measurements were performed on 3$\times$3 mm$^{2}$ size sample, the four-probe measurements were performed on 3$\times$8 mm$^{2}$ size sample with equally spaced point contacts. The electrical contacts were made using a commercial wire bonder, which utilizes aluminum metal (melted using the ultrasonic process) to make contact between the sample and the copper wires. Electrical data were also confirmed by repeating the measurements on new samples, created under identical condition, where four probe electrical contacts were made using the conventional silver paste. We also verified electrical measurements by replacing copper wire with gold wire. Similar results were obtained in all cases. 

\subsection*{Magnetic measurements} Magnetic measurements were performed in zero and applied current using a Quantum Design SQUID magnetometer with a base temperature of 5 K. Magnetic field was applied parallel to the plane of the film.

\subsection*{Spin resolved polarized neutron reflectometry measurements and DWBA modeling} Polarized neutron reflectometry (PNR) measurements were performed on a 20$\times$20 mm$^{2}$ surface area sample at the Magnetism Reflectometer, beam line BL-4A of the Spallation Neutron Source (SNS), at Oak Ridge National Laboratory. The instrument utilizes the time of flight technique in a horizontal scattering geometry with a wavelength bandwidth of 2.6 - 8.2 \AA. The beam was collimated using a set of slits before the sample and measured with a 2D position sensitive $^{3}$He detector. Beam polarization was performed using reflective super-mirror devices, achieving better than 98\% polarization efficiency over the full wavelength band. To maximize intensity, full vertical divergence was used with a 5\% $\Delta$ $\theta$/$\theta$ $\simeq$ $\Delta$ $q_z$/$q_z$ relative resolution in horizontal direction. The modeling of PNR data was performed using DWBA method, which utilizes multi-layer sample structure for scattering matrix.



\section*{Declaration of Competing Interest}
The authors declare that they have no known competing financial interests or personal relationships that could have appeared to influence the work reported in this paper.

\section*{Acknowledgments}
We thank T. Charlton and G. Vignale for help with neutron scattering experiments and theoretical discussion. DKS thankfully acknowledges the support by US Department of Energy, Office of Science, Office of Basic Energy Sciences under the grant no. DE-SC0014461. This work utilized the facilities supported by the Office of Basic Energy Sciences, US DOE.

\section*{Data availability statement}
The original data in this work is available from the corresponding author upon reasonable requests.

\section*{Appendix A. Supplementary data}
Supplementary data to this article can be found online at https://doi.org/


\begin{thebibliography}{99}


\bibitem{Zutic}  I. \v{Z}uti\'{c} et al., \text{Rev. Mod. Phys.} \textbf{76}, (2) (2004) 323-410.


\bibitem{Tulapurkar} A. A. Tulapurkar et al., \text{Nature} \textbf{438}, (2005) 339-342.

\bibitem{Kim} J. Kim et al., \text{Proc. IEEE} \textbf{103}, (1) (2015) 106-130.


\bibitem{Kent} P. Li et al., \text{Adv. Mat.} \textbf{26}, (2014) 4320.


\bibitem{Chappert} C. Chappert et al., \text{Nat. Mater.} \textbf{6}, (2007) 813-823.


\bibitem{Flatte2} D. Awschalom, M. Flatte, \text{Nat. Phys.} \textbf{3}, (2007) 153-159.


\bibitem{Dorpe} P. Van Dorpe et al., \text{Appl. Phys. Lett.} \textbf{84}, (18) (2004) 3495-3497.

\bibitem{Vignale}  M. Flatte, G. Vignale, \text{Appl. Phys. Lett.} \textbf{78}, (9) (2001) 1273-1275.

\bibitem{Acremann} Y. Acremann et al., \text{Nature} \textbf{414}, (2001) 51-54.

\bibitem{Ouna} M. Chshiev et al., \text{Europhys. Lett.} \textbf{58}, (2) (2002) 257-263.

\bibitem{Zutic2} J. Fabian et al., \text{Phys. Rev. B} \textbf{66}, (16)5301 (2002).

\bibitem{Molenkamp} A. Slobodskyy et al., \text{Phys. Rev. Lett.} \textbf{90}, (24)6601 (2003).

\bibitem{gei} G. Chern, O. Tchernyshyov, \text{Phil. Trans. Royal Soc. A} \textbf{370}, 5718-5737 (2012).


\bibitem{Glavic}  A. Glavic et al., \text{Adv. Sci.} \textbf{5}, (4) (2018) 1700856.

\bibitem{Nisoli} C. Nisoli et al., \text{Rev. Mod. Phys.} \textbf{85}, (4) (2013) 1473-1490. 

\bibitem{Stamps}  S. H. Skj{\ae}rv{\o} et al., \text{Nat. Rev. Phys.} \textbf{2}, (2020) 13-28.

\bibitem{Sondhi} C. Castelnovo et al., \text{Nature} \textbf{451}, (2008) 42-45.

\bibitem{Bramwell} S. T. Bramwell et al., \text{Nature} \textbf{461}, (2009) 956-959.



\bibitem{N} N. Rougemaille et al., \text{Phys. Rev. Lett.} \textbf{106}, (05)7209 (2011).

\bibitem{Chen2} Y. Chen et al., \text{iScience} \textbf{24}, (3) 102206 (2021).

\bibitem{Bansil} P. E. Mijnarends et al., \text{Phys. Rev. B} \textbf{65}, (07)5106 (2002).

\bibitem{Petrovykh} D. Y. Petrovykh et al., \text{Appl. Phys. Lett.} \textbf{73}, (23) (1998) 3459-3461.

\bibitem{Chern} G. W. Chern et al., \text{Phys. Rev. Lett.} \textbf{110}, (14)6602 (2013).


\bibitem{Ehlers} G. Ehlers et al., \text{J. Phys.: Condens. Matter} \textbf{16}, (11) (2004) S635-S642.

\bibitem{Lauter} V. Lauter et al., Reference Module in Materials Science and Materials Engineering (Elsevier 2016) pp. 1-27.

\bibitem{MM} R. F. L. Evans et al., \text{J. Phys.: Condens. Matter} \textbf{26}, (10)3202 (2014).

\bibitem{Zhang} S. Zhang, Z. Li, \text{Phys. Rev. Lett.} \textbf{93}, (12)7204 (2004).

\bibitem{Summers} B. Summers \text{et al.}, \text{Adv. Elec. Mat.} \textit{4}, (5) 1700500 (2018).

\bibitem{Chen} Y. Chen et al., \text{Adv. Mater.} \textbf{31}, (16) (2019) 1808298.

\bibitem{Vignale2} G. Vignale, \text{Phys. Rev. B} \textbf{71}, (12)5103 (2005).




\bibitem{Mengotti} E. Mengotti et al., \text{Nat. Phys.} \textbf{7}, (2011) 68-74.


\end{thebibliography}
\end{document}